\documentclass[conference]{IEEEtran}
\IEEEoverridecommandlockouts
\usepackage{cite}
\usepackage{amsmath,amssymb,amsfonts}
\usepackage{algorithmic}
\usepackage{graphicx}
\usepackage{textcomp}
\usepackage{supertabular,booktabs}
\usepackage{xcolor}
\usepackage[hidelinks]{hyperref}

\def\BibTeX{{\rm B\kern-.05em{\sc i\kern-.025em b}\kern-.08em
    T\kern-.1667em\lower.7ex\hbox{E}\kern-.125emX}}
\begin{document}

\title{A Threat Modelling Approach to Analyze and Mitigate Botnet Attacks in Smart Home Use Case
\thanks{[© 2021 IEEE. Personal use of this material is permitted. Permission from IEEE must be obtained for all other uses, in any current or future media, including reprinting/republishing this material for advertising or promotional purposes, creating new collective works, for resale or redistribution to servers or lists, or reuse of any copyrighted component of this work in other works.]}}

\author{\IEEEauthorblockN{Syed Ghazanfar Abbas, Shahzaib Zahid}
\IEEEauthorblockA{\textit{Al-Khawarizmi Institute of Computer} \\
Science (KICS) Lahore, Pakistan \\
ghazanfar.abbas@kics.edu.pk \\
shahzaib.zahid@kics.edu.pk}
\and
\IEEEauthorblockN{Faisal Hussain}
\IEEEauthorblockA{\textit{Al-Khawarizmi Institute of Computer} \\
Science (KICS) Lahore, Pakistan \\
faisal.hussain.engr@gmail.com}
\and
\IEEEauthorblockN{Ghalib A. Shah, Muhammad Husnain}
\IEEEauthorblockA{\textit{Al-Khawarizmi Institute of Computer} \\
Science (KICS) Lahore, Pakistan \\
ghalib@kics.edu.pk \\
muhammad.husnain@kics.edu.pk}
}
\maketitle

\begin{abstract}
Despite the surging development and utilization of IoT devices, the security of IoT devices is still in infancy. The security pitfalls of IoT devices have made it easy for hackers to take over IoT devices and use them for malicious activities like botnet attacks. With the rampant emergence of IoT devices, botnet attacks are surging. The botnet attacks are not only catastrophic for IoT device users but also for the rest of the world. Therefore, there is a crucial need to identify and mitigate the possible threats in IoT devices during the design phase. Threat modelling is a technique that is used to identify the threats in the earlier stages of the system design activity. In this paper, we propose a threat modelling approach to analyze and mitigate the botnet attacks in an IoT smart home use case. The proposed methodology identifies the development-level and application-level threats in smart home use case using STRIDE and VAST threat modelling methods. Moreover, we reticulate the identified threats with botnet attacks. Finally, we propose the mitigation techniques for all identified threats including the botnet threats.

\end{abstract}


\section{Introduction}
Internet of Things (IoT) has inaugurated the concept of enabling our daily life objects to communicate with one another with minimal human intervention to lavish human life \cite{ghazanfar2020iot}. The burgeoning applications of IoT have initiated many innovative concepts like smart home, smart city, smart parking, smart industry, smart agriculture, etc., to make the existing systems smart, intelligent, and automated. 

Despite the surging development and utilization of IoT devices, the security of IoT devices is still in infancy \cite{hossain2019application}. A recent study \cite{williams2017identifying} revealed that thousands of consumer IoT devices exposed over the internet are potentially vulnerable and most of them are webcams. The reason is that the vendors put less focus on the security of IoT devices due to race to market, race to prepare a device in less time with more features at minimum cost \cite{firdous2017modelling}. The OWASP IoT project recently reported \cite{OWASP} the top ten security flaws in IoT devices that an attacker can easily exploit to take over the IoT devices. These flaws include weak, hardcoded, or guessable passwords, lack of security updates, etc. \cite{OWASP}. The attackers first exploit these vulnerabilities, then bypass the user's privacy and information and finally use the victim IoT device to perform different malicious activities ranging from shutting down service to control over end devices \cite{xiao2019edge}. 

The rampant emergence of IoT devices caused the ignorance of security threats to large extent \cite{xiao2019edge}. The security pitfalls of IoT devices have made it easy for hackers to take over IoT devices and use them for malicious activities like botnet attacks \cite{bertino2017botnets}. Botnets are the connected network of malware-infected devices which are remotely controlled by command \& control servers \cite{xia2020modeling}. The attackers use the botnets for malicious activities like sending spam emails, click fraud, launching distributed denial of service (DDoS) attacks to chop down a web-service, etc. Botnets existed for many years, but with the proliferation of insecure IoT devices, botnets have become larger, complex and dangerous. The botnet attacks are not only catastrophic for IoT device users but also for the rest of the world since these botnets caused ever large and devastating DDoS attacks at the famous web service providers like GitHub \cite{gitAttack}, Krebs on Security, etc., in recent years \cite{xia2020modeling}. Therefore, there is a crucial need to identify the possible risks, threats, and attacks in IoT devices during the design phase so that these threats can be mitigated properly before the IoT devices are deployed. 

The above-discussed issues can be easily averted with the help of threat modelling technique. Threat modelling is a technique that is used to identify the threats in the earlier stages of the system design activity \cite{Andreas}. A threat model highlights the possible weaknesses of a system that an attacker can exploit to compromise the system \cite{Jennifer}. Based on the identified threats, different mitigation techniques are proposed in order to protect the system from the cyber-attacks. Hence, 
a threat modelling approach can be used to identify and mitigate security threats in the earlier design phase that may cause potential attacks like botnet attacks, DDoS attacks, etc. Therefore, in this paper, we propose a threat modelling approach to earlier identify the threats in a smart home use case. Particularly, we utilized two threat modelling methods, i.e., STRIDE \cite{STRIDE} and VAST \cite{VAST} in order to identify both development-level and application-level threats. Moreover, based on the identified threats, we analyzed which of the identified threats could be the reason for botnet attacks. Furthermore, we propose the mitigation strategies for all identified threats including the botnet threats to secure the underlying devices and services in a smart home use case.

\section{Literature Review}
The threat modelling is a technique that is used to identify the threats in the earlier stages of the system design activity \cite{Andreas}. There exist many threat modelling techniques like STRIDE \cite{STRIDE}, VAST \cite{VAST}, etc., that professionals and researchers use to identify the threats of a system. These techniques are summarized in \cite{shevchenko2018threat}. A threat modelling method is used to identify the potential ways that an attacker can use to compromise a system. Based on the identified threats, different mitigation strategies are proposed to protect an underlying system from the identified threats. Jennifer \textit{et al.} \cite{Jennifer} worked on the threat modelling of visual sensors network. The authors applied STRIDE-based threat modelling technique to identify possible attacks on inter-camera and intra-camera domains. Finally, they classified the identified attacks based on common weakness enumeration (CWE) and suggested some mitigation to avoid these attacks. Andreas \textit{et al.} \cite{Andreas} extended a previous work on the automation process of threat identification in software architecture using the STRIDE approach. Their proposed methodology can dynamically generate individual threats catalogues using the CWE and relate it with common vulnerabilities exposures (CVE). 

Valentina \textit{et al.} \cite{Valentina} focused on finding the data leakage threats of a case study of a home automation system in which a user interacts with the sensors in order to control the temperature and light, etc. Similarly, Laurens \textit{et al.} \cite{Laurens} proposed a threat model that provides a more holistic view of privacy standards risks by introducing the important improvements in Data Flow Diagrams (DFD).  

Matteo \textit{et al.} \cite{Matteo} focused on the threat modelling of the mobile health system. The authors used STRIDE and DREAD methodology to find out the threats and risk levels in the mobile health system. The authors also proposed some security solutions like encryption and authentication mechanisms to secure the resource-constrained gadgets.
Rafiullah \textit{et al.} \cite{Rafiullah} used the STRIDE approach to present a detailed framework for threat modelling of cyber-physical systems (CPS). They identified the possible threats and vulnerabilities in existing CPS based on the current security principles and also proposed some suggestions to mitigate the identified vulnerabilities.

The above work shows the threat modelling techniques proposed or adopted for different scenarios such as Visual Sensor Networks threat modelling, OOVL improvements, threat modelling for MicroBees, CSC threat modelling, threat modelling for the mobile health system, and CPS threat modelling. But, none of these works shows the botnet identification in their threat modelling process.

\section{Methodology}
\begin{figure}[t]
    \centering
    \includegraphics[scale=0.41]{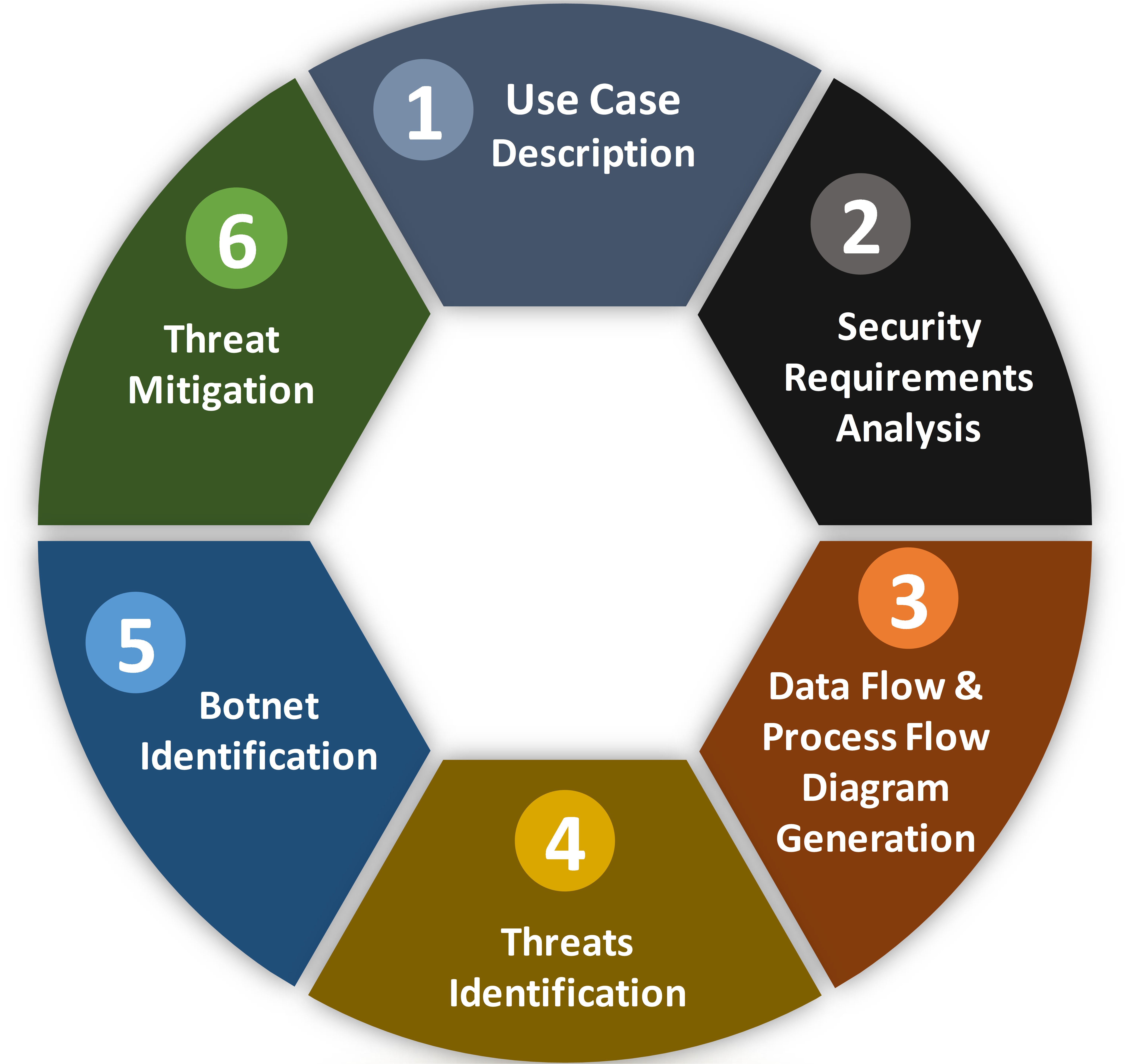}
    \caption{Proposed Threat Modeling Methodology for Smart Home Use Case}
    \label{fig:threat_model}
\end{figure}

The proposed methodology for threat modelling consists of six major steps as shown in Fig. \ref{fig:threat_model}. These steps include use case description, security requirement analysis, data flow and process flow diagram generation, threats identification, botnet identification, and threats mitigation. The detailed description of these steps is provided in the following sections.

\begin{figure*}[t]
    \centering
    \includegraphics[height=7.2cm,width=15cm]{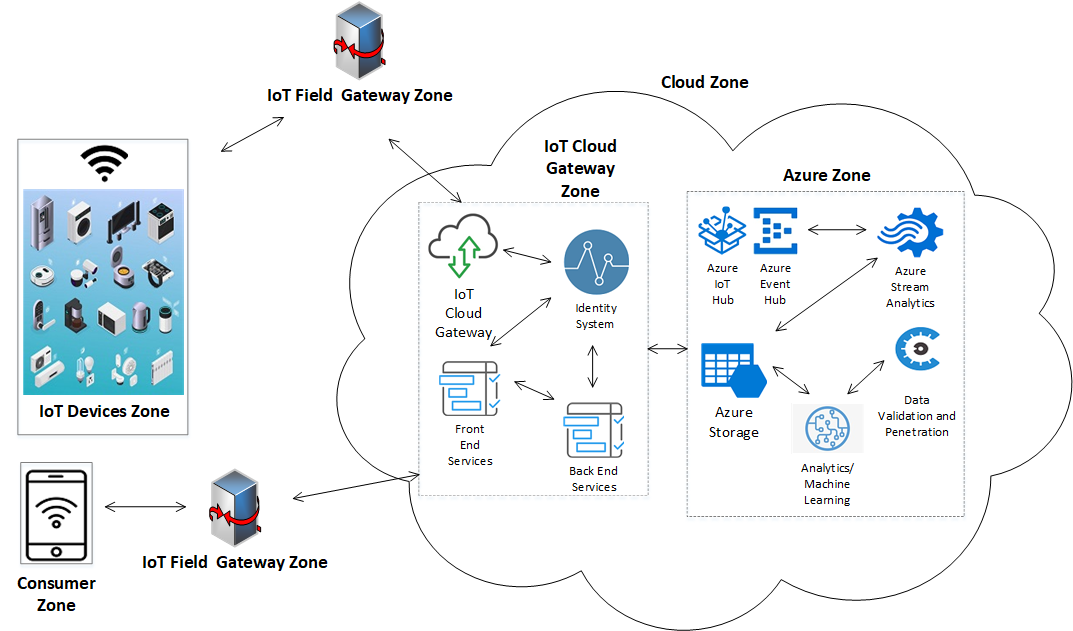}
    \caption{Smart Home Use Case Overview}
    \label{fig:azure_model}
\end{figure*}

\subsection{Use Case Descriptions}
Use case description is a primary step for threat modelling. In this paper, we considered a generic smart home use case in which any number of IoT devices can be connected with an IoT gateway and controlled via the Azure server as illustrated in Fig. \ref{fig:azure_model}. The whole use case is divided into five zones which include IoT device zone, IoT field gateway zone, Azure zone, cloud gateway zone, and consumer zone. The IoT device zone comprises of all IoT sensing and actuating devices installed in the smart home. These devices are connected to the cloud zone via the IoT field gateway as shown in Fig. \ref{fig:azure_model}. The cloud zone is the central control unit of our smart home use case. It is further divided into two sub-zones, i.e., cloud gateway zone and Azure zone. The cloud gateway zone is responsible for communication among the IoT device zone and consumer zone while the Azure zone consists of multiple components that monitor and control all IoT devices that reside in the IoT device zone. Finally, the consumer zone consists of end-user interface devices like tablets, cell phones, etc., through which a consumer can view the current status of each IoT device and also send requests to the Azure components to switch on/off IoT devices.
All the zones of our smart home use case are explained in the following sections: 

\subsubsection{IoT Device Zone}
It is a physical space having all the IoT devices of a smart home use case inside it. This zone is considered as a local zone that is separated from the public internet and may contain a short-range wireless technology through which the devices communicate with the Azure zone via IoT field gateways and cloud gateway.

The devices could be sensors like temperature sensors, humidity sensors, etc., or these could be actuators like electric water pump, electric fan, etc., or these could be embedded or computer devices like smart TV, smart lock, etc. These IoT devices could have any Linux-based operating system (OS) like RIOT, Contiki, etc., or Windows-based OS like Windows 10 Core OS, installed on them for providing interface services and controlling the execution of the services of end-device.

\subsubsection{IoT Field Gateway Zone}
Field gateway is a tool or electronic device or other general-purpose computer software that serves as a connectivity enabler, and possibly as a data processing channel for application control systems and devices. In this zone, all the IoT devices are connected to the field gateway. IoT field gateway zone could be vulnerable to physical intrusions and has limited flexibility and adaptability due to location and operational functionality constraints. 

A field gateway is distinct from a standard network router as it plays a significant role in handling connectivity and information flow among IoT devices and the Azure zone. 

\subsubsection{Cloud Zone}
This zone is divided into two sub-zones named IoT Cloud Gateway Zone and Azure Zone because both the zones operate in the cloud. The user reaches these zones through the internet and interacts with the IoT devices that are physically deployed in the home. The descriptions are as follows:
\begin{itemize}
    \item IoT Cloud Gateway Zone: Cloud gateway is responsible for the remote communication among the devices and Azure server via IoT field gateway. The cloud gateway makes the Azure server accessible for IoT device zone and consumer zone. A consumer can communicate with the Azure server from any location across the public network area. A cloud gateway can theoretically be built onto a virtualized network interface to separate all other network traffic from the cloud gateway from all of its connected devices or field gateways. In our smart home use case, this zone contains some front end and back end services that show devices information and analytics to the consumer and also transfer IoT device data and consumer requests to the Azure server. 
    
    \item Azure Zone:  This zone contains Microsoft's Azure devices. The main devices include Azure IoT Hub, Azure event hub, Azure stream analytics engine, and Azure storage as illustrated in Fig. \ref{fig:azure_model}. The Azure IoT Hub is used to connect IoT devices while the Azure event Hub is a system for the collection of sensor data at very high throughput levels from simultaneous sources. The Azure stream analytics engine enables the user to run or view real-time analytics on various data streams such as web, social media, device, sensors, etc. that store data in Azure storage. Azure zone takes input data from the IoT cloud gateway zone in such a way that the information is passed to the Azure stream analytics engine through Azure IoT Hub or Azure event Hub. This information is stored in Azure storage as recorded data which is further processed for Machine Learning analytics. 
\end{itemize}

\subsubsection{Consumer Zone}
In this zone, a service is interfaced with the user's device that is connected to the Azure server via an IoT cloud gateway. It is also responsible for data collection with the command and control process. A user can send requests over the cloud that takes the response from IoT devices.

A user can control any home device by using an Android application which is connected with the Azure server. The Azure server is the only control unit which sends commands to smart home devices for getting the sensors data or actuating the home devices. If a remote user wants to observe the room temperature or switch on/off any home appliance, he/she can initiate a request by Android application which will be forwarded to the IoT cloud gateway services. This request will be then sent to the Azure server for further processing. Now, if the user wants to see the room temperature or wants to check if the door is open or close, the Azure server will forward the current status of room temperature and door that is stored in Azure storage to the user through cloud gateway. On the other hand, if the user wants to turn on/off the fan, this request is processed by Azure stream analytics engine which will pass this request to the concerned IoT devices through Azure IoT hub via IoT field gateway and an acknowledgement is sent back to the user for the confirmation of request.

\begin{figure*}[t]
    \centering
    \includegraphics[scale=0.42]{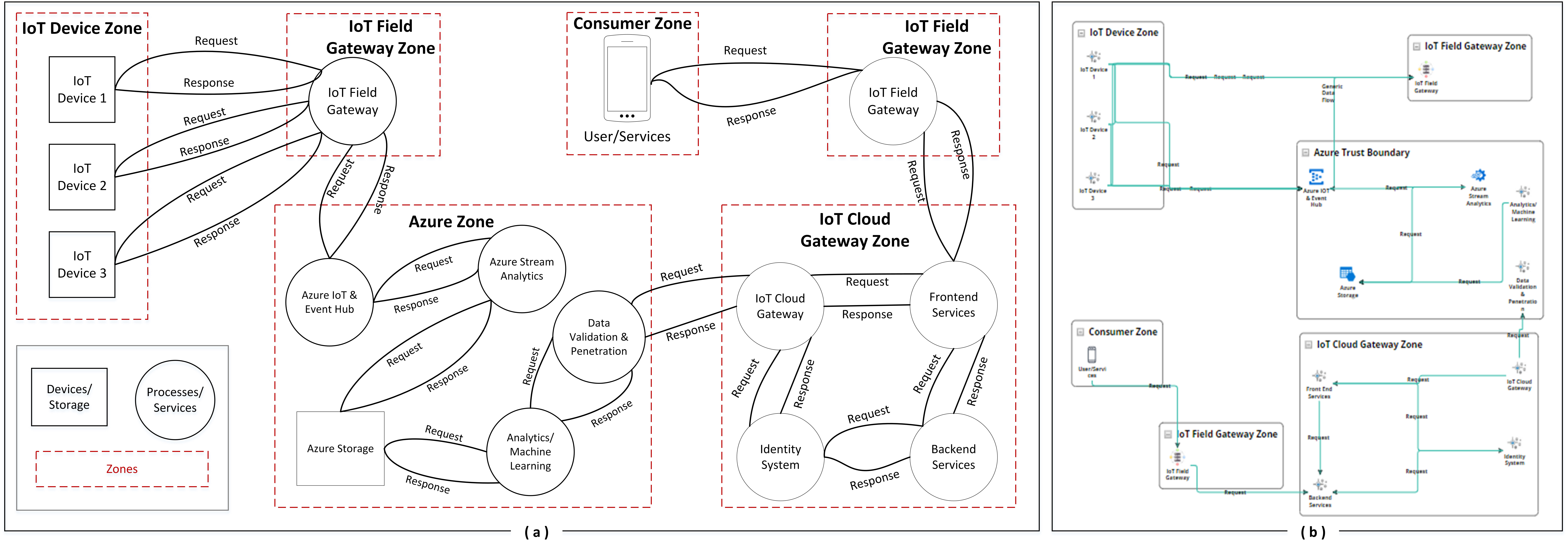}
    \caption{(a) Data Flow Diagram Designed in MTM \cite{MTM} Tool (b) Process Flow Diagram  Designed in ThreatModeler \cite{TM} Tool} 
    \label{fig:MTMmodel}
\end{figure*}

\subsection{Security Requirements Analysis}
Security requirement analysis is the foremost step in threat modelling. Here, we gather all the security requirements with respect to a use case. As mentioned earlier that in this paper, we considered a smart home use case, therefore, we gathered all the security requirements related to our use case in order to propose the threat mitigation solutions. These security requirements include implementing the authorization in IoT event hub at the protocol level, security against Denial of Services (DoS) in the device to device communication, protocol-level security on IoT field gateway, encryption on Azure storage, etc. As discussed earlier that in this work, our main focus is to prevent the smart home use case from botnet attacks, so our main goal is to propose the solution for all the identified threats that may cause botnet attacks.

\subsection{Data Flow and Process Flow Diagram Generation}
After the use case description, and the security requirements analysis, the next step is to draw the flow diagrams of the smart home use case based upon which we will identify the threats. There exists many threat modeling tools such as SecuriCAD \cite{sCAD}, ThreatModeler \cite{TM}, OWASP Threat Dragon \cite{tDragon}, IriusRisk \cite{IriusRisk}, Microsoft Threat Modeling (MTM) tool, etc. Among these, we selected the MTM \cite{MTM} and ThreatModeler \cite{TM} tools because these tools support functionality for developing an IoT use case. The MTM tool identifies the design level threats whereas the ThreatModeler \cite{TM} tool identifies the process level threats. In MTM \cite{MTM} tool we design data flow diagram (DFD) of the underlying use case whereas in ThreatModeler \cite{TM} tool we design process flow diagram (PFD) of the underlying use case. 

Fig. \ref{fig:MTMmodel}(a) shows the DFD designed for smart home use case using MTM \cite{MTM} tool. The whole diagram is designed with respect to the use case descriptions and security requirements analysis as discussed in the previous sections. In Fig. \ref{fig:MTMmodel}(a), the rectangular boxes with the black solid boundary line, represent the IoT devices and database, i.e., Azure storage while the circular shapes manifest the data processing components such as Azure stream analytics engine, front-end services, etc. The green rectangles in Fig. \ref{fig:MTMmodel}(a), represent transmission of requests/responses from one component to the other. Each zone is represented by red dotted lines. By following the previously described design hierarchy of the smart home use case, the IoT devices in Fig. \ref{fig:MTMmodel}(a) are placed on the left side and are attached to the IoT field gateway. Likewise, the IoT field gateway is connected to the cloud zone that contains two sub-zones named as IoT cloud gateway zone and Azure zone. The user in consumer zone can send requests to the Azure server that flows from the cloud zone and to the IoT device zone via IoT cloud gateway. 

As discussed earlier that the ThreatModeler \cite{TM} uses the process flow diagram (PFD) to identify the process level threats. The PFD designed using ThreatModeler \cite{TM}, is shown in Fig. \ref{fig:MTMmodel}(b). The use case descriptions and the security requirements are the same but the representation of smart home use case components in ThreatModeler \cite{TM} is more systematic as compared to the MTM \cite{MTM}. In this tool, the zones are represented by simple rectangular boxes with solid black lines. Each zone comprises of specific IoT devices/components that are mentioned in the use case descriptions. Each component is represented by its icon and green lines show the flow of requests and responses among these components.

\subsection{Threats Identification}
Once the DFD and PFD are created in MTM \cite{MTM} and ThreatModeler \cite{TM} tools respectively, the next step is to run the simulation in order to identify the threats. The MTM \cite{MTM} tool uses STRIDE \cite{STRIDE} methodology while the ThreatModeler \cite{TM} uses the VAST \cite{VAST} methodology to identify the design level and application-level threats respectively. STRIDE is the acronym for spoofing, tampering, repudiation, information disclosure, denial of service (DoS), and elevation of privileges. While VAST is the acronym for visual, agile, and simple threat modelling. 

The STRIDE \cite{STRIDE} methodology identifies the design level threat that violates any of the basic security requirements. These security requirements include confidentiality, integrity, availability, authentication, and non-repudiation \cite{STRIDE2}. Table \ref{tab:STRIDE Mapping} maps the relationship of these security requirements with the STRIDE methodology. On the other hand, the VAST \cite{VAST} methodology performs an in-depth analysis of process/application level threats. Further, it also identifies some threats that are not identified by the STRIDE methodology but it does not classify the identified threats into different categories. However, in order to better summarize the results, we classified the threats identified using VAST \cite{VAST} methodology into four categories. These four categories are mentioned and described in Table \ref{tab:VAST Mapping} based upon which the identified threats are categorized.  

\begin{table}[t]
    \centering
    \caption{STRIDE Mapping}
    \renewcommand{\arraystretch}{1.4}
    \label{tab:STRIDE Mapping}
    \begin{tabular}{|p{1.4cm}|p{4.4cm}|p{1.6cm}|}
         \hline
         \textbf{Threat Name} & \textbf{Description} & \textbf{Violation} \\
         \hline
         Spoofing & Misleading the users or systems & Authentication\\
        \hline
        Tampering & Changing the original information & Integrity\\
        \hline
        Repudiation & Denying the privileged access & Non-repudiation\\
        \hline
        Information Disclosure & Gaining the unauthorized access & Confidentiality\\
        \hline
        Denial of Services & Denying the network/system access & Availability\\
        \hline
        Elevation of Privileges & Getting the resources without the user's permission & Authorization\\
        \hline
    \end{tabular}
\end{table}

\subsection{Botnet Identification}
After the threat identification through the STRIDE \cite{STRIDE} and VAST \cite{VAST} methodologies, next comes the botnet identification step. In this step, we analyze which of the identified threats may cause a botnet attack in our smart home use case. The botnet attack occurs when an IoT device is compromised and comes under the control of bot-master. Getting a device access is the premier step for compromising an IoT device in order to make it a part of a botnet attack. In most of the recently reported botnet attacks like Mirai botnet attack, IoT devices got compromised due to the hard-coded or default passwords. The attackers compromise IoT devices either by using the default security credentials or by finding the vulnerability in the operating system (OS) or software of IoT device. In short, the attackers first get device access and then run some malware on a compromised IoT device to make it the part of the botnet. So, in our botnet identification stage, we analyze the identified threats that can affect the authentication, integrity, repudiation, and authorization of IoT devices.

\subsection{Threats Mitigation}
After the identification of potential threats in our smart home use case, the next step is to propose the mitigation techniques. Threat mitigation is a process of reducing all the possible threats in a system. In order to propose the mitigation techniques for the identified threats, we analyzed some existing 
threat mitigation studies \cite{10.1145/3376123}, \cite{anwar2020modeling}, \cite{9036313}, \cite{7876970}, \cite{7420545}. After performing the analysis on these studies, we adopt the best possible remedies in order to protect the smart home use case from the potential threats.

\begin{table}[t]
    \centering
    \caption{VAST Mapping}
    \renewcommand{\arraystretch}{1.4}
    \label{tab:VAST Mapping}
    \begin{tabular}{|p{1.5cm}|p{4.3cm}|p{1.6cm}|}
         \hline
         \textbf{Threat Name} & \textbf{Description} & \textbf{Violation} \\
         \hline
         Authentication Abuse & Threats causing exploitation of authorization or security credentials & Authentication, Authorization\\
        \hline
        Remote Code Inclusion & System and application vulnerabilities that may cause remote code execution & Integrity, Non-repudiation\\
        \hline
        Attack Hazards & Threats causing the catastrophic attacks like DoS, man in the middle, account hijacking, etc. & Availability, Confidentiality\\
        \hline
        Miscellaneous & Mixed threats including insecure update, APIs, interfaces, network services, etc. & Any\\
        \hline
    \end{tabular}
\end{table}

\section{Results and Discussion}
As discussed earlier, we performed experiments using MTM \cite{MTM} and ThreatModeler \cite{TM} tools which identify the threats using STRIDE \cite{STRIDE} and VAST \cite{VAST} methodology respectively. In our smart home use case scenario, we have five zones. We first identified the threats for each zone then analyzed the identified threats that can cause botnet attacks. As discussed earlier that both the STRIDE \cite{STRIDE} and VAST \cite{VAST} methodologies identify and map the use case threats into different categories. However, we ensembled the results of both STRIDE \cite{STRIDE} and VAST \cite{VAST} methodologies in order to better determine all the potential threats for our smart home use case. Finally, we proposed some mitigation techniques in order to secure the smart home use case from potential attacks. The detailed results are discussed in the following subsections:

\subsection{Threats Identification}
In this section we discuss all the threats identified by MTM \cite{MTM} and ThreatModeler \cite{TM} tools with respect to each zone of the use case:

\subsubsection{Device Zone Threats}
In the IoT device zone, we considered three types of IoT devices which include sensor devices, actuating devices and switching devices. Based upon the STRIDE \cite{STRIDE} and VAST \cite{VAST} methodologies, the following threats are identified in the IoT device zone.

\begin{enumerate}
    \item[i] \textbf{Spoofing Threats: } Regarding this category, Total 2 threats are reported by both MTM \cite{MTM} and ThreatModeler \cite{TM} tools in IoT device zone. These include device spoofing by reusing the authentication tokens of one device in another device, spoofing a device with a fake one and connecting to the field gateway. It will happen when the authentication token of all the devices is the same. If the adversary finds one authentication token, it can exploit all the devices at once.
    \item[ii] \textbf{Tampering Threats: } Regarding this category, Total 4 threats are reported. These include exploiting known vulnerabilities in unpatched devices, extracting cryptographic key materials, launching offline attacks by tampering the OS of devices, and intercepting encrypted traffic sent to devices. In this way, the adversary can access and change the sensitive information of devices such as sensor's values.
    \item[iii] \textbf{Authentication Abuse Threats:} Regarding this category, Total 4 threats are reported including the exploitation of unused services in devices, exploiting the permissions provisioned to the device token to gain elevated privileges, gaining unauthorized access to privileged features, and triggering unauthorized commands. In this way, the adversary has privileged access to the devices and can get full control over the devices.
\end{enumerate}

\subsubsection{IoT Field Gateway Zone Threats}
The IoT field gateway serves as an intermediate device responsible for communication among consumer, cloud and IoT devices. Following are the threats reported by both tools in the field gateway zone:
\begin{enumerate}
    \item[i] \textbf{Spoofing Threats:} Regarding this category, 3 threats are reported. These include: denying actions occur due to lack of auditing, gaining access by leveraging default login credentials, and accessing the field gateway by spoofing the IoT devices. If proper auditing is not provided on IoT field gateway, the adversary can access it with the admin rights and can mislead the current user by sending fake data of IoT devices.  
    \item[ii] \textbf{Tampering Threats:} Regarding this category, 3 threats are reported. These include getting unauthorized access, tampering the OS, and executing unknown code on the IoT field gateway. With these threats, the adversary can change the original data that flows from the IoT field gateway.
    \item[iii] \textbf{Repudiation Threats:} Regarding this category, only 1 threat is found, i.e., denying actions on field gateway due to lack of auditing which can allow the adversary to send fake data of IoT device 
    \item[iv] \textbf{Information Disclosure Threats:} Regarding this category, only 1 threat is reported, i.e., eavesdropping the communication between IoT devices and the field gateway. The adversary acts as a thief between the traffic flow of source and destination and steals personal information.
    \item[v] \textbf{Attack Hazards:} The threat here is a man in the middle attack through which the adversary can limit the availability of resources by sending the multiple requests.
    \item[vi] \textbf{Elevation of Privileges Threats:} Regarding this category, only 1 threat is reported which include getting unauthorized access to privileged features on IoT field gateway though which the adversary can monitor all the privileged resources illegally. 
\end{enumerate}

\subsubsection{IoT Cloud Gateway Zone Threats}
The following threats are reported in IoT cloud gateway zone:
\begin{enumerate}
    \item[i] \textbf{Spoofing Threats:} Regarding this category, 6 threats are reported that the adversary can exploit to spoof the user with false information.
    \item[ii] \textbf{Tampering Threats:} Regarding this category, only 1 threat is reported, i.e., tampering binaries in backend services by using tools such as IDA Pro and Sandbox. By doing this the adversary can find the strings and application programming interfaces (APIs) used by frontend and backend services.
    \item[iii] \textbf{Attack Hazards:} Regarding this category, 9 threats are reported which can allow an attacker to perform a man in the middle attack, remote code inclusion, and DoS in identity system, frontend services, and backend services.
    \item[iv] \textbf{Authentication Abuse Threats:} Regarding this category, 3  threats are reported which can affect the authorization in IoT cloud zone. 
\end{enumerate}

\subsubsection{Azure Zone Threats}
The following threats are found in Azure zone:
\begin{enumerate}
    \item[i]  \textbf{Spoofing Threats:} Regarding this category, 5 threats are reported which can allow an attacker to exploit the authentication of Azure administrator to gain access to Azure subscription.
    \item[ii]  \textbf{Tampering Threats:} Regarding this category, 6 threats are reported which can allow an adversary to eavesdrop the communication between the client and event hub and also change the integrity of data in Azure zone. 
    \item[iii]  \textbf{Repudiation Threats:} Regarding this category, 13 threats are reported that may cause account hijacking, confidential data exposure, malicious insiders, and permanent data loss in Azure IoT and event hub, Azure stream analytics. Another main threat is denying actions on Azure storage due to the lack of auditing. Hence, the adversary can steal privileged rights from the user.
    \item[iv]  \textbf{Information Disclosure Threats:}  Regarding this category, 7 threats are reported which include: abusing an insecure communication channel between a client and azure storage, weak identity, confidential data exposure in Azure IoT, event hub, Azure storage, and Azure stream analytics.
    
    \item[v]  \textbf{Attack Hazards:} Regarding this category, 6 threats are reported which can allow an attacker to perform a man in the middle attack, remote code execution to launch DoS attacks. 
    \item[vi]  \textbf{Elevation of Privileges Threats:} Regarding this category, 2 threats are reported which can allow an attacker to gain unauthorized access to resources in Azure zone and modify the data and services.
\end{enumerate}

\subsubsection{Consumer Zone Threats}
In consumer zone, two threats are reported related to the confidentiality and authorization which may cause jailbreaking into the mobile device in order to gain elevated privileges.

\begin{figure}[t]
    \centering
    \includegraphics[scale=0.41]{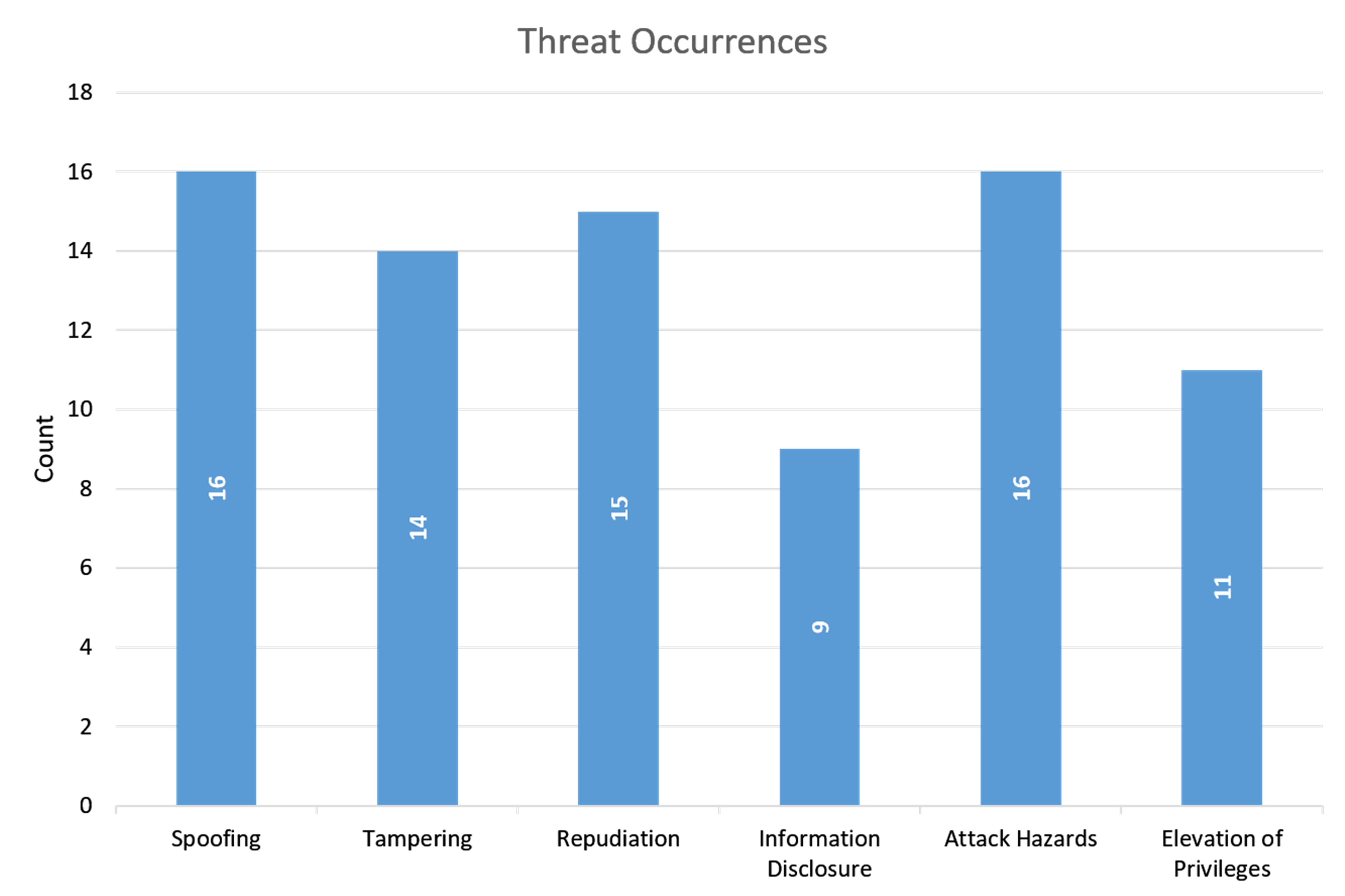}
    \caption{Overall Frequency of Threats}
    \label{fig:frequency}
\end{figure}

\subsection{Botnet Identification in Use Case}
As discussed earlier that getting a device access is the premier step for compromising an IoT device in order to make it a part of a botnet attack. Once an IoT device is compromised and comes under the control of bot-master, the attackers then use it for malicious activities like performing DoS and DDoS attacks. Fig. \ref{fig:frequency} shows the overall frequency of all the threats reported across each zone of our smart home use case. It can be observed in Fig. \ref{fig:frequency} that our smart home use case is more vulnerable to spoofing and denial of services (DoS) threats as the total number of occurrences of these threats are 16. The occurrences of tampering and repudiation are very close to these spoofing and DoS threats. 

Based upon the identified threats in the IoT device zone, an attacker can sniff the traffic coming towards the IoT devices from the IoT field gateway and get the admin interfaces or privileges. The attacker can run a malicious application or code over these compromised IoT devices, make them part of a botnet and can remotely control these devices. Likewise, in consumer zone, by utilizing the jailbreaking technique, the attacker can compromise the root of the mobile in the consumer zone and can embed or replace the malicious code with the original files and can control the user activities remotely with the root access. Similarly, due to the elevation of privileges threats reported in Azure and cloud zone, an attacker can install a malicious application or can execute the malicious code on the devices that can be remotely controlled via botmaster. In this way, all the sensitive information of the devices can be monitored and the attacker can execute its desired commands to perform unwanted actions. Therefore, if these threats are not properly managed after identification, there are high chances that the system will encounter the botnet attacks, DoS attacks, etc.

\subsection{Threat Mitigation Techniques}
Threat mitigation is a process of reducing all the possible threats in a system. Some of the threat mitigation are also found in \cite{10.1145/3376123}, \cite{anwar2020modeling}, \cite{9036313}, \cite{7876970}, \cite{7420545}. After performing the analysis on these studies, we adopt the best possible remedies in order to protect the smart home use case from the potential threats.

\subsubsection{STRIDE Threat Mitigation Techniques}
The spoofing threats can be mitigated by introducing different shared access signature (SAS) tokens and using authentication credentials on each IoT device because if the same token is used on each device, the adversary can spoof any device and become the part of the network. We can enable proper auditing in IoT field gateway using auditing rules in a firewall and by changing the default login credentials. Moreover, using Azure role-based access control (RBAC) on Azure administration, it can be guarded against spoofing. Azure storage must be protected by encryption of the information that is stored and authenticating the queries before processing. The adoption of these mitigations will not only defend against spoofing threats but also ensure confidentiality.

Similarly, by enabling the SSL/TLS communication between the client and event hub we can maintain the confidentiality and integrity data. Likewise, by enabling the UEFI secure boot and bit-locker on Windows 10 core IoT devices, we can stop the adversary to run the malicious code on devices. Additional security can be implemented by obfuscating the binaries generated during the communication between remote user and frontend or backend services in IoT cloud zone because an attacker can use various reverse engineering tools to tamper them.

\subsubsection{Botnet Threat Mitigation Techniques}
According to \cite{CRIME}, the botnet life cycle consists of five stages which include conception, recruitment, interaction, marketing, and execution (CRIME). In the conception phase, there is some kind of motivation to develop a botnet such as ego, entertainment, earn money, etc. In the recruitment phase, the botnet attack is increased to the maximum range by increasing the number of affected nodes. During the interaction phase, internal and external communication between the botnet and targeted devices is get started. After this, marketing is done by botmaster either by selling the code or by renting its services to achieve the conception. Finally, the attack is performed in the last stage. DoS is the most common attack inaugurated by botnets and in our smart home use case, DoS threat appeared most of the times. 
Therefore, our smart home use case must be protected from Dos threats and all other threats that may cause botnet attacks as discussed previously. So, we propose the following mitigation against the botnet related threats: 
\begin{enumerate}
    \item[i] \textbf{Limiting the Unused Services/Features:} An adversary can utilize the unused services or features in Azure IoT and event hub to perform the malicious actions by injecting the malicious code. These services will continue to run in the background all the time and can send significant information to the botmaster. Hence, we can limit the unused services in the cloud zone such as frontend and backend services.
    \item[ii] \textbf{Implementing Implicit Jailbreak or Rooting Detection:} By implementing the mobile root access detection mechanism can resolve the issue of jailbreaking, i.e., whenever someone wants to access the mobile root, it must be a registered user. New users must be registered via email or text message confirmation. If a user wants to log in to the system, he/she should receive a confirmation login link through email or text message.
    \item[iii] \textbf{Embedding Firewall Rules for Auditing:} Using auditing rules in a firewall at IoT field gateway and IoT cloud gateway, we can protect our smart home use case from DoS attacks. 
    \item[iv] \textbf{Encryption of the Traffic:} If the traffic flow between the user and devices is not encrypted, the adversary can monitor all the information by intercepting between the communication. However, by enabling the encryption of the traffic, we can limit the man in the middle attack.
\end{enumerate}

\section{Conclusion}
The rampant emergence of IoT devices caused the ignorance of security threats to a large extent. The attackers easily compromise these insecure IoT devices, make them a part of a botnet and use this botnet for launching devastating DDoS attacks. However, these threats can be avoided by using a threat modelling technique. A threat modelling technique can proactively identify the threat during the earlier stages of the system design activity. Therefore, in this work, we proposed a threat modelling technique for a generic smart home use case. We first identified the threats in our smart home use case using STRIDE and VAST threat modelling methods. From the identified threats, we also highlighted the certain threats that might cause the botnet attacks. Finally, we proposed mitigation strategies to alleviate the identified threats. The proposed methodology can better help the IoT vendors and developers since it provides an overview of the threats that may cause botnet attacks and also provides the mitigation strategies that can be deployed to secure the IoT devices deployed in a smart home system. This work can be further extended by implementing the proposed mitigation techniques in a real-world system to gain further insights for securing the underlying system from botnet attacks.

\bibliographystyle{IEEEtran}
\bibliography{references}

\end{document}